\documentclass[aps,superscriptaddress,showpacs,floatfix,
amsmath,amssymb,twocolumn]{revtex4}

\usepackage{epsfig}
\usepackage{dcolumn}
\usepackage{bm}

\begin{document}

\title{Spectroscopy of $^{26}$F}

\author{M.\ Stanoiu}
\affiliation{Institut de Physique Nucl\'eaire, IN2P3-CNRS, F-91406
Orsay Cedex, France} \affiliation{Grand Acc\'el\'erateur National
d'Ions Lourds (GANIL), CEA/DSM - CNRS/IN2P3, B.\ P.\ 55027, F-14076
Caen Cedex 5, France} \affiliation {IFIN-HH, P. O. Box MG-6, 76900
Bucharest-Magurele, Romania}
\author{D.\ Sohler}
\affiliation{Institute of Nuclear Research of the
Hungarian Academy of Sciences, P.O. Box 51, Debrecen, H-4001, Hungary}
\author{O.~Sorlin}
\affiliation{Grand Acc\'el\'erateur National d'Ions Lourds (GANIL),
CEA/DSM - CNRS/IN2P3, B.\ P.\ 55027, F-14076 Caen Cedex 5,
France}
\author{Zs. Dombr\'adi}
\affiliation{Institute of Nuclear Research of the Hungarian
Academy of Sciences, P.O. Box 51, Debrecen, H-4001,
Hungary}
\author{F.\ Azaiez} \affiliation{Institut de Physique
Nucl\'eaire, IN2P3-CNRS, F-91406 Orsay Cedex, France}
\author{B.A.\ Brown} \affiliation{National Superconducting
Cyclotron Laboratory, Michigan State University, East Lansing, MI 48824, USA}
\author{C.\ Borcea}
\affiliation{IFIN-HH, P. O. Box MG-6, 76900 Bucharest-Magurele, Romania}
\author{C.\ Bourgeois}
\affiliation{Institut de Physique Nucl\'eaire, IN2P3-CNRS,
F-91406 Orsay Cedex, France}
\author{Z.\ Elekes}
\affiliation{Institute of Nuclear Research of the
Hungarian Academy of Sciences, P.O. Box 51, Debrecen, H-4001, Hungary}
\author{Zs.\ F\"ul\"op}
\affiliation{Institute of Nuclear Research of the
Hungarian Academy of Sciences, P.O. Box 51, Debrecen, H-4001, Hungary}
\author{ S.\ Gr\'evy}
\affiliation{Grand Acc\'el\'erateur National d'Ions Lourds (GANIL),
CEA/DSM - CNRS/IN2P3, B.\ P.\ 55027, F-14076 Caen Cedex 5, France}
\author{D.\ Guillemaud-Mueller}
\affiliation{Institut de Physique Nucl\'eaire, IN2P3-CNRS,
F-91406 Orsay Cedex, France}
\author{F.\ Ibrahim}
\affiliation{Institut de Physique Nucl\'eaire, IN2P3-CNRS,
F-91406 Orsay Cedex, France}
\author{A.\ Kerek}
\affiliation{Royal Institute of Technology, Stockholm, Sweden}
\author{A. Krasznahorkay}
\affiliation{Institute of Nuclear Research of the
Hungarian Academy of Sciences, P.O. Box 51, Debrecen, H-4001, Hungary}
\author{M.\ Lewitowicz}
\affiliation{Grand Acc\'el\'erateur National d'Ions Lourds (GANIL),
CEA/DSM - CNRS/IN2P3, B.\ P.\ 55027, F-14076 Caen Cedex 5, France}
\author{S.M.~Lukyanov}
\affiliation{FLNR, JINR, 141980 Dubna, Moscow region, Russia}
\author{J.\ Mr\'azek}
\affiliation{Nuclear Physics Institute, AS CR, CZ 25068, Rez, Czech Republic}
\author{F.\ Negoita}
\affiliation{IFIN-HH, P. O. Box MG-6, 76900 Bucharest-Magurele, Romania}
\author{Yu.-E.\ Penionzhkevich}
\affiliation{FLNR, JINR, 141980 Dubna, Moscow region, Russia}
\author{Zs.\ Podoly\'ak}
\affiliation{Department of Physics, University of Surrey, Guildford, GU2 5XH, UK}
\author{M. G. Porquet}
\affiliation{CSNSM, CNRS/IN2P3 and Universit\'e Paris-Sud, B\^at 104-108, F-91405 Orsay, France}
\author{P.\ Roussel-Chomaz}
\affiliation{Grand Acc\'el\'erateur National d'Ions Lourds (GANIL),
CEA/DSM - CNRS/IN2P3, B.\ P.\ 55027, F-14076 Caen Cedex 5, France}
\author{M. G.\ Saint-Laurent}
\affiliation{Grand Acc\'el\'erateur National d'Ions Lourds (GANIL),
CEA/DSM - CNRS/IN2P3, B.\ P.\ 55027, F-14076 Caen Cedex 5, France}
\author{H.\ Savajols}
\affiliation{Grand Acc\'el\'erateur National d'Ions Lourds (GANIL),
CEA/DSM - CNRS/IN2P3, B.\ P.\ 55027, F-14076 Caen Cedex 5, France}
\author{G.\ Sletten}
\affiliation{Niels Bohr Institute, University of Copenhagen, Denmark}
\author{J.\ Tim\'ar}
\affiliation{Institute of Nuclear Research of the
Hungarian Academy of Sciences, P.O. Box 51, Debrecen, H-4001, Hungary}
\author{C.\ Timis}
\affiliation{IFIN-HH, P. O. Box MG-6, 76900 Bucharest-Magurele,
Romania}

\begin{abstract}
The structure of the weakly-bound $^{26}_{\;\;9}$F$_{17}$ odd-odd nucleus, produced from $^{27,28}$Na nuclei, has been investigated at GANIL by means of the in-beam $\gamma$-ray spectroscopy technique. A single $\gamma$-line is observed at 657(7)~keV in $^{26}_{9}$F which has been ascribed to the decay of the excited J=$2^+$ state to the J=1$^+$ ground state. The possible presence of intruder negative parity states in $^{26}$F is also discussed.

\end{abstract}

\date{\today}
\maketitle

The neutron-rich $^{26}_{9}$F$_{17}$ nucleus can be considered to be a \emph{benchmark} in the study of nuclear forces for at least three reasons.
Firstly, as its neutron binding
energy amounts to only 0.80(12)~MeV~\cite{Jura09} its structure is likely to be influenced by drip-line phenomena.
Secondly, it is only two nucleons off the $^{24}_{8}$O$_{16}$ doubly magic nucleus~\cite{Oza,Kanu} which has a high energy 2$^+_1$ excited state at 4.47~MeV~\cite{Hofm09}. Therefore its nuclear structure at low
excitation energy could be described by the interaction between a $d_{5/2}$ proton and a $d_{3/2}$ neutron on top of a closed $^{24}$O core, leading to $J=1,2,3,4$ positive parity states. Thirdly, it could be used to track the evolution of the negative parity states (arising from the neutron $p_{3/2}$ or $f_{7/2}$ orbits) in the $N=17$ isotones~\cite{Ober, Terr, Elek, Brown11} and predict their role in the description of the unbound $^{25}$O for which a broad resonance has been observed at 770~keV~\cite{Hofm08}.

So far the ground state spin
of $^{26}$F has been found to be $1^+$, based on the observed $\beta$-decay branches to the $0^+$ ground state
and the $2^+_1$ state in the $^{26}$Ne nucleus~\cite{Reed99}.
Two candidates of bound excited states have been proposed
at 468(17)~keV and 665(12)~keV by Elekes et al.~\cite{Elek04} with confidence levels of 2.2 and 3.8$\sigma$, respectively.
In Ref.~\cite{Elek04} the $^{26}$F nuclei were produced in interactions between a $^{27}$F nucleus and a liquid hydrogen target at 40$A$~MeV.
Moreover, a neutron unbound state has been proposed by Frank et al.~\cite{Frank11} about 270~keV above the
neutron emission threshold. This state has been populated in interactions between a $^{26}$Ne beam and
a Be target at 86$A$~MeV.

The present experimental work aims at studying bound excited states in $^{26}$F, hereby
clarifying the situation about their existence. The $^{26}$F nuclei
were produced in two-steps reactions. A primary beam of $^{36}$S with an
average intensity of 400~pnA  and an energy of 77.5$A$~MeV was used to induce
fragmentation reactions into a 398~mg/cm$^{2}$-thick C target placed
inside the SISSI device at the GANIL facility. Projectile-like fragments of interest were
selected through the $\alpha$ spectrometer. A wedge-shaped,
130~mg/cm$^{2}$-thick, Al foil was installed at the dispersive focal
plane between the two dipoles of the spectrometer to provide an
additional energy-loss selection. The magnetic rigidity of the
$\alpha$ spectrometer was optimized for the transmission of
secondary beam nuclei with energies of about 60$A$~MeV, among which $^{27,28}$Na
were the main genitors of $^{26}$F. An 'active' target composed of
a plastic scintillator (103.5 mg/cm$^{2}$) sandwiched between two
carbon foils of 51 mg/cm$^{2}$ was used for identifying the
nuclei of this cocktail beam through their time of flight values.
This 'active' target was also used to induce secondary reactions.
The $^{26}$F nuclei were subsequently selected and identified on an event-by-event basis through the SPEG spectrometer using the time of flight, energy loss, and focal-plane position
information. The $^{26}$F nuclei could be produced in the ground state, as well as in bound or unbound
excited states. The short-lived ($<1~ns$) excited bound states decay by prompt in-flight
$\gamma$ emission, while (if existing) longer-lived or unbound states could not be observed in our work. To detect prompt $\gamma$-rays, the 'active' target was surrounded by an array of 74 BaF$_2$ detectors located in two hemispheres at
a mean distance of 30~cm from the target. Doppler shift corrections were applied to
the observed $\gamma$-rays as a function of their detection angle to account for the momentum value of the emitting nuclei determined
by the measurement of their position at the focal plane of SPEG located 22 meters downstream to the BaF$_2$ array. The total photo-peak efficiency of the BaF$_2$ array was about 42\% (29\%) for a $\gamma$-ray of 600~keV (1300~keV). The energy resolution ($\sigma$), including the part
provided by the Doppler broadening, amounts to about 10\% of the $\gamma$ energy at 500~keV. Low-energy $\gamma$ transitions down to 100~keV were detected with an
efficiency value of about 24\%.

\begin{figure}
\centering \epsfig{height=11cm,file=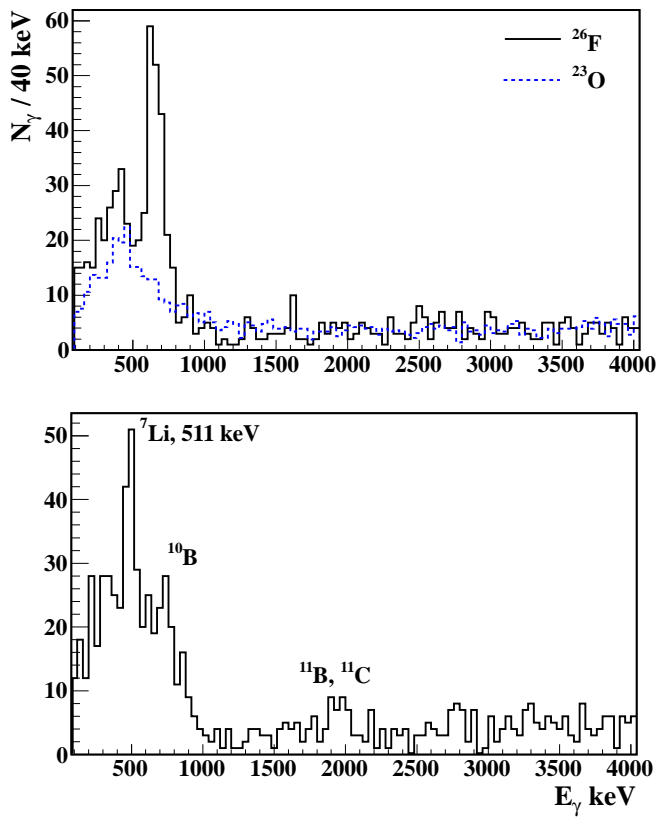} \caption{Top: Doppler
corrected $\gamma$-ray spectrum obtained using the BaF$_2$ detectors for $^{26}$F (full line), and for $^{23}$O (dotted line). Bottom: Spectrum obtained for $^{26}$F without applying the Doppler corrections. Labeled $\gamma$-rays belong to target fragmentation.}
\label{26Fspectrum}
\end{figure}

The Doppler-corrected $\gamma$-spectrum obtained for $^{26}$F is
shown in the upper part of Fig.~\ref{26Fspectrum}. It exhibits a
clear peak at 657(7)keV, with a width $\sigma$=61(6)keV. The centroid energy of the peak matches, within the experimental value, the one tentatively
proposed by Elekes et al.\cite{Elek04}
at 665(12)~keV. The low-energy structure seen around 400~keV (with $\sigma$=81(14)keV) is
too broad to correspond to a $\gamma$-ray emitted in flight. Indeed, known in-flight $\gamma$-peaks observed at a similar energy during the present experiment have smaller widths: the $\gamma$ peak at 478~keV in $^{7}$Li has $\sigma$=51(2)keV while that at 320~keV in $^{11}$Be has $\sigma$=36(1)keV.
Moreover the same pattern appears around 400~keV in the $^{23}$O nucleus (see dashed line in upper part of Fig.~\ref{26Fspectrum}), which has no bound excited state~\cite{Stan04}. Therefore this structure more likely corresponds
to a $\gamma$-transition arising from the fragmentation of the target nuclei. As these $\gamma$-rays
are  emitted almost at rest, they are not suitably Doppler-corrected when using the
velocity of the in-flight nuclei, leading to the broad structure in the $\gamma$-ray spectrum (upper panel of Fig.\ref{26Fspectrum}).
Without applying any Doppler correction a thinner peak,
belonging to target excitations, appears at about 500~keV in Fig.~\ref{26Fspectrum} (lower panel)  on top of a broader structure, caused by the 657~keV $\gamma$-rays emitted in flight. The
tentative $\gamma$-ray  at 468(17)~keV proposed by Elekes et al.\cite{Elek04} may  either be hidden in the background of our spectrum, or not be fed
by the presently used reaction.

The structure of $^{26}_9$F$_{17}$ can be viewed simply by assuming a closed
$^{24}$O core to which a proton and a neutron are added. The
$^{24}$O nucleus has the properties of a doubly magic nucleus~\cite{Oza,Kanu}, with a first
excited state above 4.47~MeV~\cite{Hofm09} and a neutron $N=16$ shell gap of about 4.95(16)~MeV.
Following the normal filling of orbits, the odd valence proton  occupies the $d_{5/2}$ orbit, while the
odd valence neutron occupies the $d_{3/2}$ one. Their coupling leads to states
$J=1^+-4^+$. Guided by the rule of particle-particle couplings for a neutron and a proton having the same $\ell$ value on top of a
closed core (see for instance ~\cite{Cast}), a parabolic curve as a function of $J$
is expected to be formed by the
states in $^{26}$F, the lowest spin $J=1$ and highest spin  $J=4$ spin states having the
largest binding energy, i.e. lowest excitation energy. This is what is found by the Shell Model calculations
presented in Table~\ref{TableI}. The calculated configuration of
these states is as much as 80-90\% pure $\pi d_{5/2} \otimes \nu
d_{3/2}$. Noteworthy is the fact that both the USD~\cite{usd} and USDA/USDB~\cite{usdab}
interactions predict the $J=4$ state to be a $\beta$-decaying isomeric state,
partly connected to the ground state by a delayed $M3$ transition.
In all interactions, the $J=3$ state is found to be unbound,
at an excitation energy of about 1.7~MeV. Taking the value of 0.80(12)~MeV for the neutron
emission threshold, the $J=3$ state is predicted to be unbound by about 0.9~MeV. Consequently, it is reasonable to discard the two possibilities of $J=4$ and $J=3$ for the observed excited state at 657(7)~keV, the former being possibly long-lived isomer, the latter being likely unbound. We therefore ascribe the observed peak at 657(7)keV to arise from the decay of the $J=2$ excited state to the $J=1$ ground state.

Besides the 'normal' positive parity states,
low-lying negative parity states could be present in $^{26}_{9}$F$_{17}$. Indeed a 3/2$^-$ intruder state has been discovered at 765~keV above the 3/2$^+$ ground state in the
$^{27}_{10}$Ne$_{17}$ isotone~\cite{Ober,Terr,Elek}. From the recent work of Ref.\cite{Brown11},
it is confirmed that the 3/2$^-$ state has a large $p_{3/2}$ component. Therefore the coupling of the
$p_{3/2}$ neutron to the $d_{5/2}$ proton would lead to negative parity states $J=1^--4^-$ in $^{26}$F.
The excitation energy of these negative parity states should
be related to the energy difference between the neutron $d_{3/2}$ and $p_{3/2}$ orbits, which is
expected to be reduced in $^{26}$F as compared to $^{27}$Ne.
This reduction comes from the fact that the removal of a proton from the $d_{5/2}$ orbit between $^{27}$Ne and $^{26}$F should
weaken the binding energy of the  $\nu d_{3/2}$ orbit relative to the $\nu p_{3/2}$ one. This owes to the
fact that the $\pi 1d_{5/2} - \nu 1d_{3/2}$ monopole interaction (between orbits having the same number of nodes $n$,
the same angular momentum $\ell$ and anti-aligned spin) is expected to be more attractive than the
$\pi 1d_{5/2} - \nu 2p_{3/2}$ one (between orbits with different $n$ and $\ell$ values, and aligned spin)~\cite{Sorl08}.
It follows that a multiplet composed of negative parity states should be present at relatively low excitation
energy in $^{26}$F. The Hamiltonian WBP-M used in the $0s-0p-0d-1s-0f-1p$ valence space in Ref.\cite{Brown11} predicts the lowest negative parity states in $^{26}$F to be $2^-$ and $4^-$ around 1~MeV excitation. In order to produce these negative parity states, the projectile nuclei should contain a significant intruder configuration component (such as $(p_{3/2})^2$) added to the dominant $(d_{3/2})^2$ one. This is not likely in the present experiment, in which the $^{26}$F nuclei were produced from $^{27,28}$Na, which lie outside of the island of inversion according to Ref.\cite{Trip} and Fig. 9 of Ref.\cite{usdab}. Energy-wise, the observed unbound state by Frank et al.~\cite{Frank11} at about 270~keV above the 0.8~MeV neutron emission threshold
would correspond to a $3^+$ or $2^-,4^-$ state. However it is again hard to conceive a favored feeding of negative parity states from a $^{26}$Ne precursor nucleus, which is expected to contain a negligible fraction of $(2p_{3/2})^2$ component in its ground state. The 468(17)~keV $\gamma$-ray,
\emph{tentatively} observed by Elekes et al.\cite{Elek04} through the $^{27}$F(-1n)$^{26}$F reaction,
might correspond to the decay of a negative parity excited state. Indeed, hint for a significant intruder content of the ground state of $^{27}$F was derived from the large deformation length in the (p,p') reaction~\cite{Elek04}.

\begin{table}
\caption{Comparison between experimental and calculated energies in keV of the $1^+$ to $4^+$ states in $^{26}$F.
Calculations are made with shell model calculations using the USD~\cite{usd}, USDA or USDB~\cite{usdab} interactions.
$^*$ Energy extracted from Ref.~\cite{Frank11}, however the spin assignment is not known (see text for details)}
\label{TableI}
\begin{ruledtabular}
\begin{tabular}{lllll}\noalign{\smallskip}
$J$& $1$&$4$ & $2$& $3$ \\
\noalign{\smallskip}\hline\noalign{\smallskip}
Exp& 0 & &657(7) & $\simeq$~1070(120)$^*$\\
USD&0&353&681&1604\\
USDA& 0 & 850&850&1800\\
USDB& 0&200&600&1600\\
\noalign{\smallskip}
\end{tabular}
\end{ruledtabular}
\end{table}

To summarize, the $^{26}$F nuclei were produced in two-steps fragmentation reactions. Their in-flight
$\gamma$-decay were observed in a large efficiency detector array composed of 74 BaF$_2$ detectors. A single
bound state is proposed from the observation of a 657(7)~keV $\gamma$-ray.  Among the possible J=$1^+-4^+$ spin
values of the low energy states in $^{26}$F predicted by shell model calculations, we propose a
spin assignment $J=2^+$ for this state. It is based on the facts that the $J=4^+$ state is likely to be an
isomer, and that the $J=3^+$ state is predicted to lie above the neutron emission threshold.
Further experimental investigations are needed to locate the $J=4^+$ state and to conclude about
the presence of negative parity (intruder) states in $^{26}$F. The presence of negative parity states at low excitation energy would suggest a further reduction of the neutron $d_{3/2} - p_{3/2}$ spacing in $^{26}$F, as compared to $^{27}$Ne. If pursued, this reduction would manifest itself as a $p_{3/2}$ component at low-energy in the $^{25}$O isotone. To evidence the presence of $\ell=1$ negative parity states in the $^{26}$F and $^{25}$O nuclei, further experiments should be carried out using the fragmentation of projectile nuclei (such as $^{31}$Na) lying inside the island of inversion.


\acknowledgments {\small This work has been supported by the
European Community contract N$^{\rm o}$ RII3-CT-2004-506065, by
OTKA K68801 and K100835, by the NSF PHY-1068217 grants, by a grant of the Romanian National Authority for Scientific Research, CNCS - UEFISCDI, project number PN-II-ID-PCE-2011-3-0487 and the Bolyai J\'anos Foundation. J.C. Thomas is acknowledged for comments on the manuscript.}



\begin{thebibliography}{0}
\expandafter\ifx\csname natexlab\endcsname\relax\def\natexlab#1{#1}\fi
\expandafter\ifx\csname bibnamefont\endcsname\relax
  \def\bibnamefont#1{#1}\fi
\expandafter\ifx\csname bibfnamefont\endcsname\relax
  \def\bibfnamefont#1{#1}\fi
\expandafter\ifx\csname citenamefont\endcsname\relax
  \def\citenamefont#1{#1}\fi
\expandafter\ifx\csname url\endcsname\relax
  \def\url#1{\texttt{#1}}\fi
\expandafter\ifx\csname urlprefix\endcsname\relax\def\urlprefix{URL }\fi
\providecommand{\bibinfo}[2]{#2}
\providecommand{\eprint}[2][]{\url{#2}}

\end{thebibliography}


\begin{thebibliography}{100}


\bibitem{Jura09} B. Jurado et al. Phys. Lett. B 649,43(2007).
\bibitem{Oza} A. Ozawa et al., Phys. Rev. Lett. 84, 5493 (2000).
\bibitem{Kanu} R. Kanungo et al., Phys. Rev. Lett. 102, 152501 (2009).
\bibitem{Hofm09} C. R. Hoffman et al. Phys. Lett. B  672, 17 (2009).
\bibitem{Ober} A. Obertelli, et al., Phys. Lett. B 633, 33 (2006).
\bibitem{Terr} J. Terry, et al., Phys. Lett. B 640, 86 (2006).
\bibitem{Elek} Z. Elekes, et al., Phys. Rev. Lett. 96, 182501 (2006).
\bibitem{Brown11} S. M. Brown et al, submitted to Phys. Rev. C
\bibitem{Hofm08} C. R. Hoffman et al., Phys. Rev. Lett. 100, 152502 (2008).
\bibitem{Reed99} A. T. Reed et al., Phys. Rev. C 60, 024311 (1999).
\bibitem{Elek04} Z. Elekes et al. Phys. Let. B 599, 17 (2004).
\bibitem{Frank11} N. Frank et al., Phys. Rev. C 84, 037302 (2011).
\bibitem{Stan04} M. Stanoiu et al. Phys. Rev. C 69, 034312 (2004).
\bibitem{Cast} R. F. Casten, Nuclear structure from a simple perspective, chapter 4, Oxford university press.
\bibitem{usd} B. A. Brown and B. H. Wildenthal, Ann. Rev. of Nucl. Part. Sci. 38,  29 (1988).
\bibitem{usdab} B. A. Brown and W. A. Richter, Phys. Rev. C 74,  034315 (2006).
\bibitem{Sorl08} O. Sorlin and M. G. Porquet, Prog. Part. Nucl. Phys. 61, 602 (2008).
\bibitem{Trip} V. Tripathi et al., Phys. Rev. C 73, 054303 (2006).









\end{thebibliography}
\end{document}